\documentclass{aa}  
\bibpunct{(}{)}{;}{a}{}{,} % to follow the A&A style
\usepackage{graphicx}
\usepackage{xcolor}
\usepackage{txfonts}
\usepackage{hyperref}
\hypersetup{colorlinks = true, linkcolor=red, citecolor=blue, urlcolor=blue}
\usepackage{natbib}
\begin{document} 
   \def\ergs{erg\,s$^{-1}$}
   \def\msunyr{M$_{\odot}$\,yr$^{-1}$}
   \def\kms{\mbox{km~s$^{-1}$}}
   \def\mgiidoub{\ion{Mg}{ii}\,$\lambda\lambda$\,$2795,\,2802$}
   % \title{Using CSST to constrain the ejecta-CSM interaction in Type II supernovae and their progenitors}
   \title{Using CSST and ejecta-wind interaction in type II-P supernovae to constrain the wind-mass loss of red supergiant stars}
   \titlerunning{Ejecta-wind interaction in SNe II with CSST}
   \authorrunning{J. Luo et al.}

   \author{Jingxiao Luo\inst{1,2,3,4}
          \and
          Luc Dessart\inst{5}
          \and
         Xuefei Chen\inst{1,2,4}
          \and
          Zhengwei Liu\inst{1,2,4}
          }

   \institute{Yunnan Observatories, Chinese Academy of Sciences (CAS), Kunming 650216, P.R. China\\
              \email{zwliu@ynao.ac.cn, cxf@ynao.ac.cn}
         \and
             Key Laboratory for the Structure and Evolution of Celestial Objects, CAS, Kunming 650216, P.R. China
        \and
            University of Chinese Academy of Sciences, Beijing 100049, P.R. China
        \and
            International Centre of Supernovae, Yunnan Key Laboratory, Kunming 650216, P.R. China
        \and
            Institut d’Astrophysique de Paris, CNRS-Sorbonne Université, 98 bis boulevard Arago, F-75014 Paris, France
             }

   \date{Received xxx xx, 202x; accepted May 30th, 2024}

% \abstract{}{}{}{}{} 
% 5 {} token are mandatory
 
  \abstract
  {The properties of H-rich, type II-plateau supernova (SN II-P) progenitors remain uncertain, and this is primarily due to the complexities associated with red supergiant (RSG) wind-mass loss. Recent studies have suggested that the interaction of the ejecta with a standard RSG wind should produce unambiguous signatures in the optical (e.g., a broad, boxy H$\alpha$ profile) and in the UV (especially Ly\,$\alpha$ and \mgiidoub) a few years following the explosion. Such features are expected to be generic in all SNe II-P and can be utilized to constrain RSG winds. Here, we investigate the possibility of detecting late-time (0.3-10 years since explosion) SNe II-P in the NUV with the China Space Station Telescope (CSST). Convolving the existing model spectra of ejecta-wind interactions in SNe II-P with the transmission functions of the CSST, we calculated the associated multiband light curves, in particular, the NUV (255\,nm -- 317\,nm) band, as well as the $NUV-r$ color. We find that the CSST will be able to detect the NUV radiation associated with ejecta-wind interaction for hundreds SNe II-P out to a few hundred Mpc over its ten-year main sky survey. The CSST will therefore provide a sizable sample of SNe II-P with the NUV signatures of ejecta-wind interaction. This will be helpful for understanding the mass loss history of SN II-P progenitors and their origins.}


   \maketitle
%
%________________________________________________________________

\section{Introduction}

Massive stars with initial masses above $\mathrm{}8\,M_{\sun}$ are typically believed to end their lives as core collapse supernovae (CCSNe) and may have exert a variety of influences on their surroundings and host galaxies.
These SNe may contribute to the production of heavy elements, trigger new star formation and play a role in stellar feedback in galaxy formation and evolution \citep[see][]{matteucci1986relative, nomoto2006nucleosynthesis,2013ARA&A..51..457N,2014MNRAS.445..581H,gnedin2016physical}.
However, there are still open questions about how these massive stars evolve to the point of collapse and explode \citep[see][for reviews]{2012ARA&A..50..107L,janka2012explosion, burrows2021core}.
A Better understanding of CCSNe may shed light on how these massive stars evolve and, more interestingly, what happens in their cores shortly before their collapse.

In this work, we focus on SNe II, which originate from the collapse of the core of massive stars and show hydrogen lines in their spectra \citep[see][for reviews]{1997ARA&A..35..309F, turatto2003classification, 2017hsn..book..195G}.
They contribute to about half of the observed SN demography \citep[in a magnitude limited sample, see][for details]{2011MNRAS.412.1522S, 2011MNRAS.412.1441L, perley2020zwicky}.
In recent decades, various efforts have been made to connect SNe II to their respective progenitors.
Red supergiants (RSGs) have been established as progenitors for the most common H-rich SNe \citep[SNe II-P or SNe II-L, see][]{2004ApJ...617.1233Y, 2007ApJ...661.1013L, 2011MNRAS.410.1739D, 2011MNRAS.412.1522S}.
However, the observed progenitor sample for SNe II-P and L seems to have a upper mass cutoff at around $\mathrm{}17\,M_{\sun}$, while stellar evolution models commonly predict the upper mass cutoff to be $25\sim30\,M_{\sun}$ \citep[see][for details]{2009ARA&A..47...63S, 2015PASA...32...16S}.
% Therefore this apparent lack of SNe II-P/L progenitors in the mass range of $17\sim25\,M_{\sun}$ has been coined the \textit{``red supergiant problem''}, which is an unsolved question related to SNe II and their progenitors.

Various methods have been proposed to constrain CCSNe progenitors:
(i) direct searches for progenitors that rely heavily on serendipitous data from the Hubble Space Telescope (HST). This HST progenitor sample size grows at the speed of about one or two new progenitor verification per year \citep[see][]{2018MNRAS.474.2116D};
(ii) shock breakout and cooling modeling of early light curve of SNe can be utilized to constrain some properties of the progenitor at the time of explosion \citep[see][for a review]{Waxman2017};
(iii) Another way to constrain the properties of the progenitors is to look at the circumstellar material (CSM), which is formed by mass loss events during the life of the progenitor \citep{smith2014mass,2017hsn..book..403S}. 
Some types of CCSNe, such as SNe IIn (e.g., SN 2010jl; \citealt{2011ApJ...730...34S}) and SNe Ibn and Icn (e.g., SN 2006jc;  \citealt{2007ApJ...657L.105F, 2008MNRAS.389..113P, sun2020origins}, and SN 2021csp; \citealt{2022ApJ...927..180P}, also see \citealt{2022A&A...658A.130D}), require a period of enhanced mass loss from the progenitor star just prior to the explosion to explain their behaviour (e.g., \citealt{2006A&A...460L...5K, 2007ApJ...656..372G}).
(iv) Also, modeling the SN spectra (with e.g., \textit{SYNAPPS}, see \citealt{2011PASP..123..237T}) and fitting the SN light curve can be utilized to derive the basic properties of the SN ejecta, such as the total ejecta mass, ejecta velocity, and amount of $^{56}$Ni produced in the explosion \citep[see][for examples]{1989ApJ...340..396A,1993A&A...273..106B,1994ApJ...429..300W,2019A&A...631A...8H,2019ApJ...879....3G}. This information can be further used to constrain the properties of the progenitor star. However, this method may suffer from degeneracies among parameters \citep[see][for details]{2019A&A...625A...9D}. %Modelling of spectra and light curve

Compared to other types of CCSNe, as well as SNe IIn and Ibn and, Icn are relatively rare in number \citep{2009ARA&A..47...63S, perley2020zwicky} and they exhibit diverse signatures due to interaction between the SN ejecta and CSM.
The strength of the interaction signatures depends on the CSM mass, which range from $\mathrm{0.1}\,M_{\sun}$ to a few solar masses and even up to tens of solar masses in extreme cases \citep[see][]{2012AJ....144..131Z, 2012ApJ...757..178G}.
For the SNe IIn with higher inferred CSM masses (for example in SN 2010jl; \citealt{2014ApJ...797..118F, dessart2015numerical}), the light curve modeling often requires outburst-like events with mass loss rates on the order of $\mathrm{0.1}\,M_{\sun}$\,yr$^{-1}$ or higher.
In the case of SNe Ibn, there are evidences that the enhanced mass loss could be a natural result of massive helium star evolution \citep[see][for details]{2022A&A...658A.130D}.
These pre-SN eruptions or violent mass loss episodes form a CSM that is dense and extended enough to significantly boost the SN optical luminosity and alter its light curve (see \citealt{2020A&A...637A..73N} for an analysis on a sample of 42 SNe IIn, and \citealt{2014AJ....147..118R} for a comparison between peak magnitude distribution of different types of SNe).
For SNe II with dense CSM material close to the surface of the progenitor, the ejecta-CSM interaction may still induce noticeable changes, especially in the form of early phase flash-ionization features \citep{yaron2017confined,2017A&A...605A..83D,2023arXiv231114409L}, but the effect on the overall light curve may be marginal beyond a few weeks\citep{2023arXiv231016885I}.

While the enhanced mass loss events prior to explosion of some SNe are of great interest \citep{2023ApJ...952..119B}, an universal feature of all massive stars is to experience a steady wind-mass loss throughout their post-main sequence evolution \citep[see][]{maeder1987grids, langer1994towards, smith2014mass}.
While the wind velocities are typicaly high in blue supergiants and Wolf-Rayet stars, the low wind velocity in RSG stars implies a high wind density.
As RSGs explode as SNe II-P/L, the wind material produced by steady-state mass loss during the $\mathrm{10^6\,yrs}$ RSG phase will fill the SN vicinity.
This circumstellar wind material, albeit less dense than CSM formed by eruptive events, is expected to take thousands of years to be fully swept up by the SN ejecta
Therefore the interaction between the ejecta and the wind material will affect the evolution of SNe and their supernova remnants (SNRs) on the timescales of centuries.
Detecting this circumstellar wind material in nearby SNe II-P and inferring the long-term pre-SN mass loss rate \citep{2023MNRAS.523.1474R} may possibly improve our understanding of massive star evolution and help in constraining current stellar evolution models.

Past analyses \citep{2003LNP...598..171C} and recent numerical simulations \citep{dessart2022modeling} show that interaction between SNe II ejecta and an optically-thin CSM produced by wind-mass loss inject a substantial shock power into the ejecta. 
Based on these results, for a steady mass loss rate on the order of $\mathrm{10}^{-5}M_{\sun}\,\mathrm{yr}^{-1}$, which is common during massive star evolution and detected around many supergiants \citep[see][]{langer1994towards, wang2021red}, the corresponding shock power will surpass radioactive heating from $\mathrm{0.1}M_{\sun}$ of $^{56}$Ni about one year after the SN explosion.
For a lower mass loss rate of $\mathrm{10}^{-6}M_{\sun}\,\mathrm{yr}^{-1}$, the shock power surpasses decay later at about 700\,d.
One of the effects of this interaction has been observed as broad boxy H$\alpha$ emission several years after explosion (e.g., in spectra of SN 2004et three years after explosion; \citealt{2012PASP..124..668Y}, also in SN 2014G; \citealt{2016MNRAS.462..137T}, in SN 2017eaw, SN 2017ivv, and various other SNe; see \cite{2023A&A...675A..33D} for additional examples).
From the spectra and analysis by \citet{dessart2022modeling} and \citet{2023A&A...675A..33D} we know that a significant part of the shock power injected into the SN ejecta is absorbed.
The temperature of the outer ejecta can reach over 15\,000\,K, thus re-emits the energy it absorbs as UV radiation, while leaving the optical luminosity of the SN almost unaffected. 
Only in the $U$ band and at shorter wavelengths does the effect of additional shock power become obvious. 
In their Pwr1e41 model, the bulk of late-time UV radiation introduced by SN ejecta-wind interaction is released as strong Ly$\alpha$ and \mgiidoub\ line emissions.

Unfortunately, the \mgiidoub\ doublet is not observable from the ground due to heavy atmospheric absorption.
Some rare exceptions are the  superluminous SNe detected at high redshifts and in which the rest-frame NUV lines are observed on Earth in the optical (e.g., iPTF16eh, a SLSN at z=0.427, \citealp[see][]{lunnan2018uv}). 
Therefore, to observe and study the UV features of nearby SNe, a space-based telescope is required.
Past HST observations have shown that some SNe exhibit late-time UV emission up to 1000 days after explosion (e.g., SN2017eaw, \citealp[see][]{van2023disappearances}), which can be explained by late-time SN ejecta-wind interaction \citep[see][]{2020ApJ...900...11W}.
However, it takes a long time for the HST-based sample size to grow considerably.
Meanwhile, the existing UV SN light curves from swift UVOT are typically limited to SNe brighter than 20 mag, \citep[see][]{brown2009ultraviolet}; thus they are not deep enough for late-time monitoring of SNe in the UV.
Ideally, a space-based deep UV survey with large FOV is more suitable for identifying a large number of year-old SNe II-P.

Here, we examine the possibility of utilizing the upcoming China Space Station Telescope (CSST) \citep[see][for an overview]{zhan2018overview} to survey and identify SN ejecta-wind interaction signatures in nearby SNe over a large sky area. 
One of the main missions of CSST is to conduct a multi-band sky survey over its designed operation lifetime of ten years \citep{zhan2021wide}.
The multiband sky survey of CSST includes NUV-band (255nm-317nm), in which the \mgiidoub\ doublet from SN ejecta-wind interaction will land.
Photometry from other bands and low resolution spectra from gratings of CSST will also provide additional information on SNe it observes.
Previously, \citet{2022A&A...666L..14D} showed that broadband photometry of nebular phase SNe can be possibly utilized as a tool to constrain their progenitors, as only few emission lines remain dominant at this stage and they fall into separate photometry filters.

\begin{table*}
	\centering
	\caption{Magnitude limits, wavelength range, and spacial resolution of the CSST sky survey.}
%	Remember to define the quantities, symbols and units used.}
	\label{tab:csst_performance}
	\begin{tabular}{lccccccc} % four columns, alignment for each
		\hline
		 & NUV & u & g & r & i & z & Y\\
		\hline
		Wavelength Range (nm) & 255--317 & 322--396 & 403--545 & 554--684 & 695--833 & 846--1000 & 937--1000\\
		Limiting Magnitude & 25.4 & 25.4 & 26.3 & 26.0 & 25.9 & 25.2 & 24.4\\
		Spacial Resolution (") & 0.135 & 0.135 & 0.135 & 0.135 & 0.145 & 0.165 & 0.165\\
		\hline
	\end{tabular}
	\\[10pt]
    {\raggedright \textbf{Note}: Data from \citet{zhan2021wide}. Total exposure times are 600s for NUV, 300s for u/g/r/i/z and 600s for Y. Magnitude limits are values when expected signal-noise ratio is equal to 5. \par}
\end{table*}

Here, we investigate the detectability of ejecta-wind interaction in SNe II-P using the NUV filter onboard CSST, including the impact of distance and redshift, the specific survey procedure, and limited spectroscopic classifications.
We surmise that CSST will be able to provide a sizeable sample for SNe II-P ejecta-wind interaction from its main survey alone.
We look forward to the CSST data serving as a basis for constraining RSG mass loss and improve the mass loss recipes currently used on stellar evolution modeling.

The rest of this paper is organized as follows. 
We first discuss the expected photometric performance of the CSST survey system and the input spectral templates of SNe II with late-time ejecta-wind interaction in Section~\ref{sec:methods}.
We then calculated the simulated SN magnitudes and colors observed by CSST in Section~\ref{sec:results}.
A simple strategy combining ground-based data with the CSST survey data to identify SN with wind interaction is presented in Section~\ref{sec:strat}. 
Section~\ref{sec:conclusion} concludes this paper.
% In the following parts of this paper, "ejecta" refers to supernova ejecta, and "ejecta-CSM interaction" refers to interaction between the supernova ejecta and the supernovae CSM unless specified.
All magnitudes presented in this paper are in the AB system. We adopted the $\Lambda$CDM cosmology with $H_{0}$ = 70 km s$^{-1}$ Mpc$^{-1}$, $\Omega_{\Lambda}$ = 0.7, and $\Omega_{\text{m}}$ = 0.3.

\section{Methods and inputs}
\label{sec:methods} % used for referring to this section from elsewhere
In this section, we make a brief introduction of CSST, its relevant survey mode and characteristics of its onboard instrument. 
We also briefly describe the SN ejecta-wind interaction models we used as input.
We then convolved the filter transmission functions of the CSST sky survey with the modeled spectra of ejecta-wind interaction.
This gives the expected observed magnitudes of ejecta-wind models by CSST, which we further discuss in Section~\ref{sec:results}.
%Normally the next section describes the techniques the authors used.
%It is frequently split into subsections, such as Section~\ref{sec:maths} below.

\subsection{CSST expected performance}
\label{sec:performance}
China Space Station Telescope is a 2-m aperture space telescope expected to be launch-ready by the end of 2024. 
It will operate at a low-Earth Orbit with a height of around 400 km, sharing the same orbit as China's Tiangong Space Station.
During the ten-year-long main operation phase of CSST, 70\% of the telescope time will be allocated to conduct the main sky survey covering a sky area of 17500 deg$^2$ \citep{zhan2018overview, zhan2021wide}.

The main sky survey of CSST will utilize a wide FOV survey camera that consists of 30 9K × 9K CCDs in a 5 × 6 array.
The spatial-resolution of the CSST sky survey is expected to be 0.135" to 0.165", depending on wavelength.
Each CCD sensor on the survey camera is covered by a broadband filter (four NUV and Y filters and two filters each for ugriz, with 18 in total) or grating (four gratings each for GI, GV, GU, with 12 in total) and has a FOV of 11' by 11' each.
In combination, the CSST survey camera will have a total FOV of 1.1 deg$^2$.

When the main sky survey is performed, CSST will conduct photometric observation of each field within the 17500 deg$^2$ footprint either two or four times (two for ugriz bands, and four for NUV/Y bands and gratings). 
In addition, a deeper survey covering 400 deg$^2$ will be carried out.
The limiting magnitudes and other characteristics of the CSST sky survey \citep[see][for details]{zhan2021wide} is summarized in Table~\ref{tab:csst_performance}.

The wavelength range available for the CSST sky survey is 255nm-1000nm. 
Specifically, the 255nm-317nm NUV band will be very useful to identify the \mgiidoub\ emission feature, while other bands might be utilized to better understand the local environment of the SN.

\begin{figure*}
	% To include a figure from a file named example.*
	% Allowable file formats are eps or ps if compiling using latex
	% or pdf, png, jpg if compiling using pdflatex
	\includegraphics[width=2.0\columnwidth]{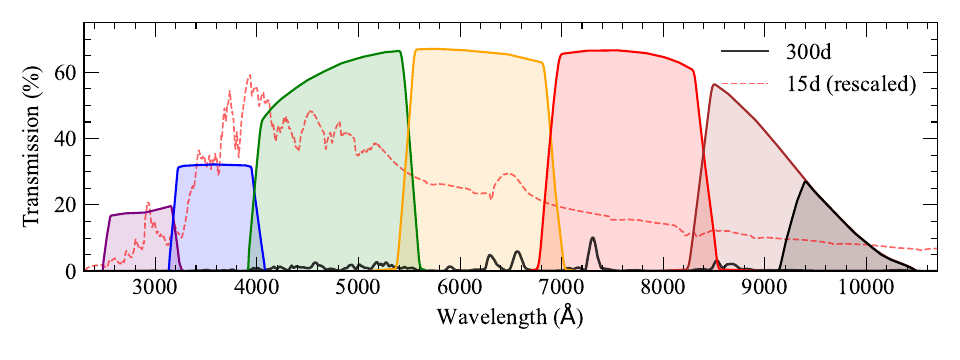}
    \caption{Normalized spectra of model Pwr1e41 from \citet{dessart2022modeling} plotted over the filter transmission function of the CSST sky survey. The purple line and shaded region under it denotes the transmission function of NUV band. The red dashed line is the SN spectrum 15 days after explosion (re-scaled by an factor of 0.3), when most of power is released in optical range. The solid black line is the model spectrum at 300 days after explosion, when reprocessed shock power dominates the UV luminosity in the form of Ly$\alpha$ (falling in the FUV and thus not shown) and the \ion{Mg}{ii} $\lambda\lambda$ 2795, 2802 doublet in NUV band. The other shaded regions are transmission functions of u/g/r/i/z/Y bands.}
    \label{fig:CSST_filters}
\end{figure*}

%Simple mathematics can be inserted into the flow of the text e.g. $2\times3=6$
%or $v=220$\,km\,s$^{-1}$, but more complicated expressions should be entered
%as a numbered equation:

%\begin{equation}
%    x=\frac{-b\pm\sqrt{b^2-4ac}}{2a}.
%	\label{eq:quadratic}
%\end{equation}

%Refer back to them as e.g. %equation~(\ref{eq:quadratic}).

\subsection{Interaction models}
In this study, we used the SN II-P models with shock interaction and their spectral time sequences from \citet{dessart2022modeling}. All SN models arise from a solar metallicity star with an initial mass of $\mathrm{15}M_{\sun}$, then evolved and exploded as a typical SN II-P with an ejecta mass of $\mathrm{10.81}M_{\sun}$ \citep[see][for details]{2019A&A...631A...8H}. 
The total kinetic energy of the ejecta is 1.3\,×\,$10^{51}$\,erg with 0.03M$_{\sun}$ of $^{56}$Ni initially.
The ejecta density profile is then modified to have a $\mathrm{0.1}M_{\sun}$ outermost dense shell travelling at a speed of $V_{\rm sh}$ = 11700\,km\,s${-1}$.

After explosion, these ejecta are then assumed to encounter material formed by the RSG progenitor steady-state wind-mass loss.
The ejecta interacts with the wind material and the shock power is deposited back into the ejecta's dense shell as X-rays, which is then absorbed and processed by the ejecta in the same way as radioactive decay radiation.

Different amounts of interaction power are injected continuously and at steady rates into different models, ranging from zero to 10$^{43}$\,erg\,s$^{-1}$.
This injected power range corresponds to a pre-SN mass loss rate from zero to 10$^{-3}$M$_{\odot}$\,yr$^{-1}$, under the assumption of a) The CSM is formed by a steady stellar wind with velocity V$_{\infty}$=50\,km\,s${-1}$; b) hydrodynamic interaction between the CSM and the ejecta doesn't change the ejecta's density and velocity profiles significantly; c) all of the shock power is absorbed by the ejecta.
In total, eight models with different interaction powers were generated (zero, 10$^{40}$, 10$^{41}$, 5\,×\,10$^{41}$, 10$^{42}$, 2.5\,×\,10$^{42}$, 5\,×\,10$^{42}$, and 10$^{43}$\,erg\,s$^{-1}$). Eight nonlocal thermodynamic equilibrium, time-dependent radiative-transfer simulations were presented in \citet{dessart2022modeling}.
The model with no interaction power peaks around -17.4\,mag in absolute magnitude in g and r band of CSST, and has a optically thick duration (i.e., plateau) of around 100 days, which reproduces the observational characteristics of typical SNe II-P \citep{2014ApJ...786...67A}.

Three out of eight spectral sequences were modeled to $\mathrm{t}$\,=\,300\,d since explosion (zero, 10$^{40}$, 10$^{41}$ erg s$^{-1}$, named NoPwr, Pwr1e40 and Pwr1e41, respectively), other sequences terminate earlier, but show similar trends and evolve more slowly. 
For our study, we mainly focused on these three models as they require the least amount of pre-SN mass loss (which means that their progenitors do not need an enhanced period of mass loss prior to the SN explosion, thus might be more common). 
Also, CSST is expected to visit SN sites long after the explosion, therefore the $\mathrm{t}$\,=\,300\,d models are more representative, although we note that the NUV interaction feature is expected to last much longer \citep[see][]{2023A&A...675A..33D}.
For interaction models with higher interaction power, they share a similar evolution trend with the three models we discuss here, but with even an higher UV luminosity.

As discussed in previous sections, the majority of the interaction power emerges in the UV, especially in the form of Ly$\alpha$ and \mgiidoub\ emission lines.
While Ly$\alpha$ is out of reach for CSST unless at high redshift, the NUV band of the CSST main sky survey is suitable for capturing the \ion{Mg}{ii} $\lambda\lambda$ 2795, 2802 signal from nearby (z<0.14) SNe.
In Fig.~\ref{fig:CSST_filters}, the $\mathrm{t}$\,=\,300\,d spectra of Pwr1e41 is plotted together with the normalized CSST sky survey camera transmission functions.

It is clear from Fig.~\ref{fig:CSST_filters} that for this model, apart from Ly$\alpha$ (not shown in the figure), the majority of the $\mathrm{t}$\,=\,300\,d UVOIR radiation emitted by the SN is concentrated in the \ion{Mg}{ii} doublet, and the NUV band happens to capture this spectral feature very well, while other bands receive a small fraction of that NUV flux.
This suggests that the SNe influenced by such ejecta-wind interaction will be bright in NUV band, especially when compared with other optical/IR bands.

In order to study the long-term CSST-NUV brightness evolution of ejecta-wind interaction, we further incorporated spectra from \citet{2023A&A...675A..33D}, whereby two SN II models (s15p2NoPwr and s15p2Pwr1e40) were evolved to 1000 days after explosion. 
These two models start from the s15p2 model presented in \citet{2021A&A...652A..64D}.
The s15p2NoPwr model was evolved without ejecta-wind interaction power and the s15p2Pwr1e40 was evolved with an extra 10$^{40}$\,erg\,s$^{-1}$ of ejecta-wind interaction power, using the same approach at that described in \citet{dessart2022modeling} but for an ejecta mixed with a shuffled-shell technique \citep{2021A&A...652A..64D}.

\section{Results}
\label{sec:results}

In this section, we present the results from convolving the synthetic spectra with the CSST filter transmission functions. We also discuss the photometric properties of ejecta-wind interaction in SNe II.
\subsection{Magnitude evolution}
\begin{figure}
	% To include a figure from a file named example.*
	% Allowable file formats are eps or ps if compiling using latex
	% or pdf, png, jpg if compiling using pdflatex
	\includegraphics[width=\columnwidth]{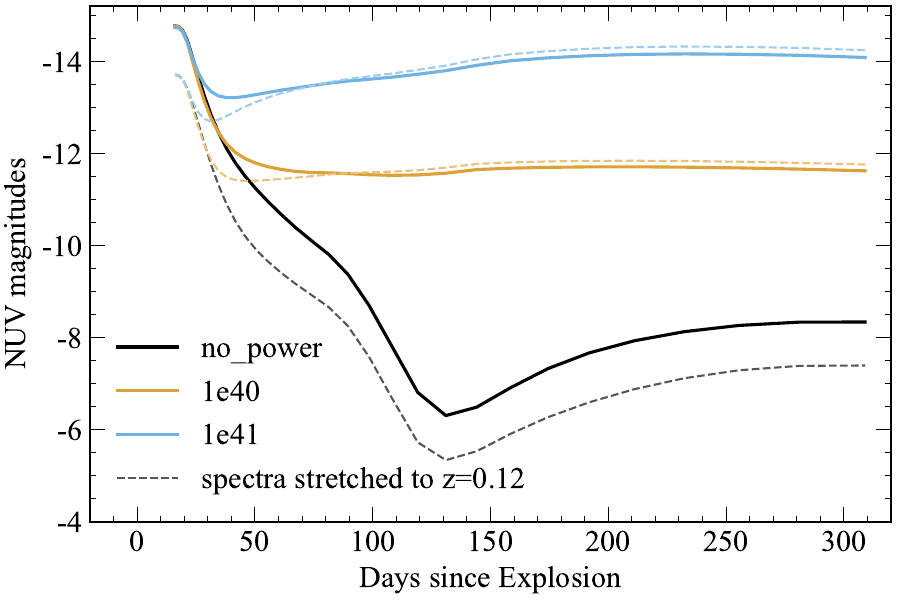}
    \caption{NUV magnitude evolution of nopwr, Pwr1e40 and Pwr1e41 models of \citet{dessart2022modeling}. The thick lines represent light curves at z=0 in absolute magnitude. The dashed thin lines represent magnitudes of the models after application of a wavelength stretch equivalent to a redshift of 0.12, which brings the \mgiidoub\ lines to the red edge of NUV filter range.}
    \label{fig:CSST_mags}
\end{figure}

The thick lines in Fig.~\ref{fig:CSST_mags} show the CSST-NUV light curves in absolute magnitude for the three models we selected (i.e., models with zero, 10$^{40}$ and 10$^{41}$ erg s$^{-1}$ of injected power, corresponds to an pre-SN mass loss rate of zero, 10$^{-6}$ and 10$^{-5}$M$_{\sun}$\,yr$^{-1}$). 
We note that based on direct observations of its environment, Betelgeuse ($\alpha$ \textit{Orionis}) has an average mass loss rate of about 2 × 10$^{-6}$\,\msunyr \citep[see][for a summary of this star]{dolan2016aOri}.
Another study conducted by \citet{wang2021red} shows RSGs in M31 and M33 exhibit an average mass loss rate of 2 × 10$^{-5}$\,\msunyr.
Furthermore, based on the analysis of late-time broad boxy H$\alpha$ emission caused by ejecta-wind interaction, the progenitors of SN 2004et and SN 2017eaw are estimated to have experienced a mass loss rate of 2$\sim$3 × 10$^{-6}$\,\msunyr \citep[see][for details]{2009ApJ...704..306K, 2020ApJ...900...11W}.
These previous results suggest that for our choice of models, the respective progenitor mass loss rates are typical and representative.  

The effect of ejecta-wind interaction is very prominent in the NUV band, even with weak interaction power. 
The models with interaction power start to diverge from the zero power model in less than 50 days from explosion.
As the zero-power reference model fades in the NUV, the models with interaction power stay around a sizable NUV luminosity for an extended period of time.
During this time, the models with interaction power can be a few magnitudes brighter than the reference model, which allows the NUV part of the SN to remain detectable for a much longer period of time over a larger distance.
This phenomenon was noticeable in the UVW2 band light curves originally calculated in \citet{dessart2022modeling} (see their Fig. 2), with similar trends and differences in magnitude.

In the case of distant SNe at higher redshifts, the \mgiidoub\ feature is shifted to longer wavelengths, it will eventually fall out of the NUV band and begin to land in the u band.
We examine this effect by stretching the spectra of our models to mimic the impact of redshift at z=0.12, and the result is displayed as the dashed thin lines in Fig.~\ref{fig:CSST_mags}.
It shows that even at z=0.12, the overall trend of the light curves does not change significantly.
This is because that the shift in wavelength is partly compensated by the higher efficiency of the sensor at the red edge of the NUV band.
The redshift of z=0.12 is around the furthest detection range within the scope of this paper, as the plan is to identify NUV-bright SNe candidates based on ground-based SNe discovery data, which is incomplete beyond the redshift of z=0.12.

\subsection{NUV color evolution}
\begin{figure}
	% To include a figure from a file named example.*
	% Allowable file formats are eps or ps if compiling using latex
	% or pdf, png, jpg if compiling using pdflatex
	\includegraphics[width=\columnwidth]{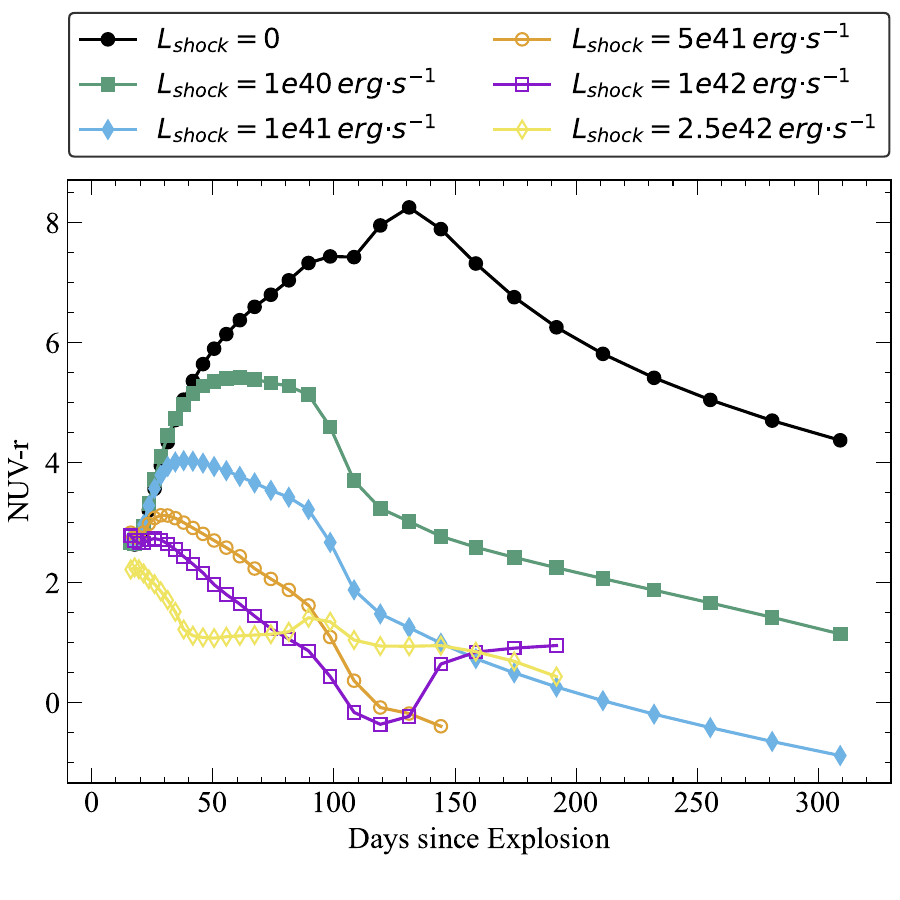}
    \caption{NUV-r color evolution of different ejecta-wind interaction models of \citet{dessart2022modeling}. Each line corresponds to a different ejecta-wind interaction power injected into the model, denoted as different values of $\mathrm{L_{shock}}$. The models with higher interaction power exhibit significantly bluer color.}
    \label{fig:CSST_color}
\end{figure}

Apart from the persistent NUV flux, another signature of ejecta-wind interaction is the distinctive color evolution.
As long as the fading optical flux from the SN remains detectable (either from ground-based telescopes or by CSST itself), the color difference between the NUV and optical bands will be strongly noticeable as shown in Fig.~\ref{fig:CSST_color}. 
A similar trend is observed in NUV-g and NUV-i colors (see Appendix for plots of other color combinations).

In general, the NUV-g/r/i colors of models with low or no interaction power become redder after the explosion as the ejecta cool and fade.
However, the NUV-g/r/i colors turn blue after a certain amount of time (for example at t\,=\,130 days for model nopwr), this reflects the transition into the nebular phase.
The color behaviour in this phase is largely determined by the relative fluxes and wavelengths of various emission lines, and no longer directly reflects the temperature of the ejecta.
The models with interaction energy experience this transition earlier, as the \mgiidoub\ emission from ejecta-wind interaction significantly boost the NUV luminosity, while the optical lines continuous to fade.
Overall, the models with interaction power exhibits significant bluer NUV-g/r/i colors than the non-interaction model.

This difference in NUV-g/r/i colors might be helpful to better differentiate ejecta-wind interaction events from other types of long-lasting transients, such as SLSNe, as our light curves predict sources radiating predominately in the UV range.
It may also be possible to identify and categorize previously unknown SNe with strong interactions based on their color; however, this is beyond the scope of the present study.

\subsection{Detection range}
\label{sec:detection}
Based on design specification from \citet{zhan2021wide} (listed in Table~\ref{tab:csst_performance}), the expected limiting magnitude of the NUV filter for the CSST sky survey is m$_{NUV}$\,=\,25.4\,mag. 
However this is the cumulative limiting magnitude from a total of four 150s exposures gathered between two visits, which might be separated by several years.
Two consecutive 150\,s NUV exposures are expected from a single scan by CSST, which will result in a slightly shallower limiting magnitude.
Using data from a single scan is suitable for our study since they are only separated by 30\,s of slewing and re-centering, reflecting a single epoch of SN evolution.

\begin{table}
	\centering
	\caption{Absolute NUV magnitudes at t=300d of different models observed in the CSST-NUV band, and the expected detection distance (redshifts) by the CSST sky survey.}
%	Remember to define the quantities, symbols and units used.}
	\label{tab:csst_detection}
	\begin{tabular}{cccc} % four columns, alignment for each
		\hline
		Interaction & $\mathrm{M_{NUV}}$ & Detection & Redshift\\
		power & t=300d & distance & \\
		\hline
		None & -8.0\,mag & <30 Mpc & \\
		10$^{40}$erg/s & -11.5\,mag & <100 Mpc & <0.025\\
		10$^{41}$erg/s & -14.0\,mag & <400 Mpc & <0.1\\
		\hline
	\end{tabular}
\end{table}

\begin{figure}
	% To include a figure from a file named example.*
	% Allowable file formats are eps or ps if compiling using latex
	% or pdf, png, jpg if compiling using pdflatex
	\includegraphics[width=\columnwidth]{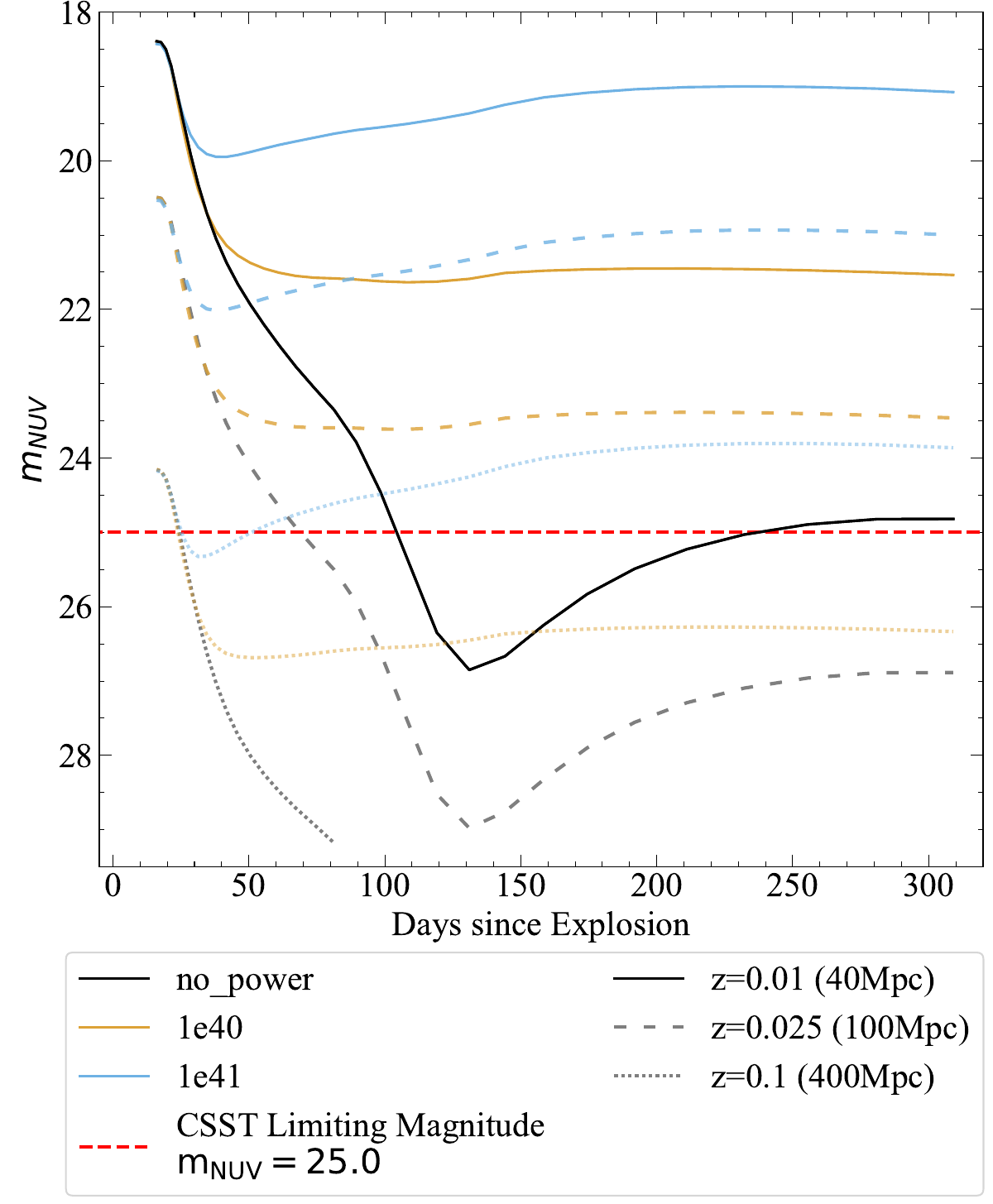}
    \caption{Apparent NUV magnitude evolution of model nopwr, Pwr1e40 and Pwr1e41 of \citet{dessart2022modeling}. The solid lines represent light curves at a distance of z=0.01 ($\sim$\,40\,Mpc), the dashed thin lines represent light curves at z=0.025 ($\sim$\,100\,Mpc) and the dotted lines represent light curves at z=0.1 ($\sim$\,400\,Mpc). The red dashed thick line represents the limiting magnitude of a 2×150 second exposure in NUV band for CSST.}
    \label{fig:CSST_appr_mags}
\end{figure}

Therefore, we decided to take m$_{NUV}$\,=\,25.0\,mag for our criteria of limiting magnitude for a single CSST visit.
We estimated the detection range of ejecta-wind interaction SNe II based on this limiting magnitude.
For the three models discussed above, the absolute magnitude and respective detection range is listed in Table~\ref{tab:csst_detection}, their light curves in apparent magnitude is displayed in Fig.~\ref{fig:CSST_appr_mags}.

For the model with 10$^{41}$ erg s$^{-1}$ of interaction power, it will remain detectable for CSST at a distance of 400 Mpc ($\sim$z=0.1) for at least a year after its explosion (likely remains bright for a decade, inferred from the evolution trend of its light curve). 
At this distance and redshift, the \mgiidoub\ feature remains well inside the CSST-NUV band; hence, using NUV brightness as a proxy for ejecta-wind interaction is appropriate throughout the detection range of CSST.

As the CSST sky survey will likely be the first deep, mostly unbiased and high angular resolution NUV survey to be carried out, we expect to detect NUV signatures from a large number of SNe II-P within nearby galaxies. 
In the case where the mass loss rate of the progenitor RSG wind is high or eruptive, the resulting NUV signature is expected to be brighter and can be detected by CSST from a greater distance.
Based on our model and with no additional input power, NUV photometry from a non-interacting SNe II-P will remain detectable for a year after explosion out to 30\,Mpc for CSST, likely together with emission in other filter bands.

This distance-limited (less than 30\,Mpc) sample of SNe II-P can potentially serve as a critical diagnostic for stellar evolution modeling and the properties of the pre-SN star, as they may vary in NUV brightness, which reflects different amounts of mass loss of their RSG progenitors.
For SNe II-P from a typical RSG progenitor that has experienced typical mass loss, our models show that ejecta-wind interaction would cause it to be bright in NUV for at least a few years.
A non-detection or a very dim SNe II-P in NUV would indicate a very small mass loss rate in the RSG phase (10$^{-7}$\,\msunyr or lower) of its progenitor.
This would imply that the progenitor evolves without any mass loss and therefore dies with a mass that is essentially the same as its initial mass.

For SNe II-P at further distances (30 to 400 Mpc), the CSST data may miss some of the NUV-faint SNe and their emissions in the optical bands are unlikely to be detectable for CSST at late-time.
However the total number of detected SNe II-P in the NUV band may be far greater than the 30 Mpc sample due to the much larger volume; thus, it may provide a sizable sample for future statistical study.
\subsection{Long-term evolution}
\label{sec:long}
As mentioned in Section~\ref{sec:performance}, the CSST main sky survey is planned to last ten years, visiting every part of its footprint twice.
At this rate, it is unlikely for CSST to image SNe shortly after their explosions.
Instead, most of the observed SNe II will be those that have faded for a few years.
Therefore, it's important to know how these ejecta-wind interaction signatures fade on longer timescales.

\begin{figure}
	\includegraphics[width=\columnwidth]{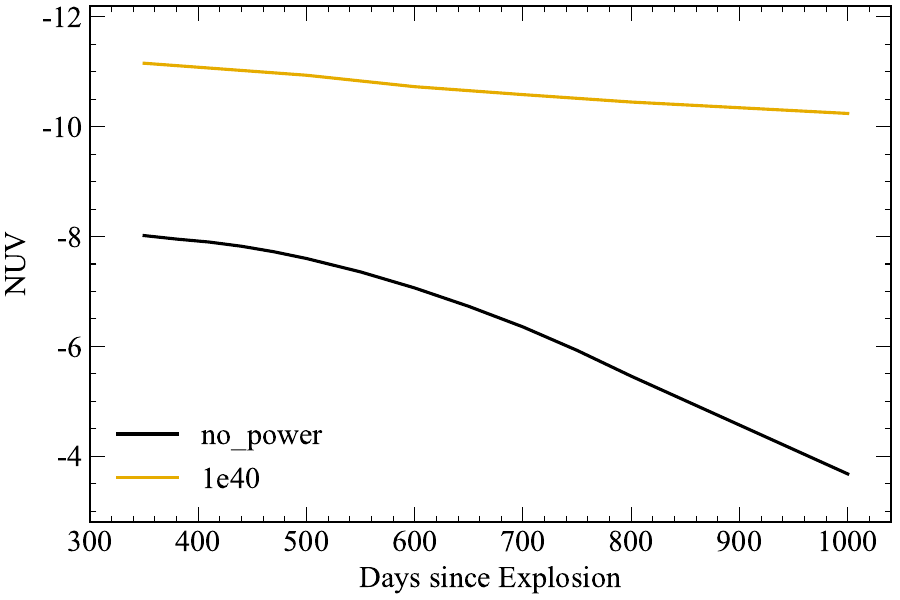}
    \caption{Absolute NUV magnitude evolution of s15p2NoPwr (black), s15p2Pwr1e40 (yellow) model of \citet{2023A&A...675A..33D}, starting from 350 days since explosion to 1000 days since explosion.}
    \label{fig:late_mags}
\end{figure}

In this section, we used the model spectra from \citet{2023A&A...675A..33D} and calculated their CSST-NUV band brightness from t=350d to t=1000d, the result is shown in Fig.~\ref{fig:late_mags}.
For the model without additional power input, the NUV absolute magnitude exhibits a drop of around four magnitudes, or about two mag per year, effectively rendering all non-local-group SNe of this type undetectable three years after explosion.
However, for the model with the most modest amount of ejecta-wind interaction power input (just 10$^{40}$\,erg/s), its NUV brightness is predicted to drop more slowly, by about 0.9 mag from t=350d to t=1000d.
This corresponds to a NUV decline rate of about 0.5 mag per year, keeping the ejecta-wind interaction signatures in NUV band visible for much longer.
For example, in the range of 30\,Mpc, model with 10$^{40}$\,erg/s of power input is expected to remain visible for nearly a decade after exploding (fading from M$_{NUV}$\,=\,-11.5\,mag at T\,=\,300\,d to M$_{NUV}$\,$\sim$\,-8\,mag at T\,=\,8\,yrs after the explosion).

The slow decline rate is mainly due to the fact that the power released from SN ejecta impacting the wind remains constant, and starts to slowly overtake radioactive decay as the predominant power source in the SN.
The predicted decline is mainly driven by the density drop of the SN ejecta as it expands, the dominant emission line shift from \mgiidoub\ to Ly$\alpha$ and other lines.
Combined with exponential decline of power released by radioactive decay, the flux observed in the NUV band is expected to decrease slightly over time.

With this rate of decline, the ejecta-wind interaction signatures are expected to last for at least a few years, even possibly out to a decade.
Therefore CSST will have ample time to survey the sky and detect long-lasting NUV flux from at least some of the older SNe that have exploded some years prior.
Judging by the velocity ratio between the ejecta and the RSG wind ($V_{sh}$/$V_{\infty}$ is about 200), the interaction between the ejecta and the wind material can potentially lasts for centuries, as some RSGs spend at least $\mathrm{10^6}$\,yrs in this phase.
During this time the interaction power stays high and the NUV flux may last for much longer.
The predicted long-lasting NUV flux, along with radio and X-ray observations can bridge the gap between SNe and young SNRs, a transition phase which is currently not fully explored as these old SNe and young SNRs are often optically faint \citep[see][for some historical SNe that have transitioned into young SNRs]{2017hsn..book.2211M, 2020ApJ...890...15F,van2023disappearances,2023MNRAS.523.1474R}.

\subsection{Comparisons with existing examples}
\label{sec:comp}
\subsubsection{SN 2017eaw}
SN 2017eaw in NGC 6946 is known for showing signs of late-time ejecta-CSM interaction \citep{2020ApJ...900...11W, van2023disappearances}, as it has a F275W magnitude of 22.7 about 3.5 years after explosion (as well as 24.1, 23.8, and 23.1 mag in F336W, F555W, and F814W, respectively).
Unfortunately, there is no late-time UV spectrum of this object which covers the NUV wavelength range (255\,nm -- 317\,nm) of CSST. 
Therefore we cannot directly reconstruct the expected CSST NUV magnitudes from existing observational data.

Based on photometic measurements alone, it is evident that SN 2017eaw would have been detectable by CSST 3.5 years after explosion at various bands, as the limiting magnitudes of CSST are around 25 mag.
This is not unexpected since SN 2017eaw was relatively close (about 7 Mpc), in spite of the significant galactic and host extinction ($E(B\,-\,V)_{gal}$ = 0.30 mag and $E(B\,-\,V)_{tot}$ = 0.41 mag according to \citet{2019ApJ...876...19S}).
If SN 2017eaw were free from these extinction effects, its NUV magnitudes would have been $\sim\,2$ mags brighter, assuming the extinction law from \citet{1999PASP..111...63F} and $R_V$ = 3.1.
\subsubsection{SN 2010jl and SN 2005ip}
These two SNe are both classified as SNe IIn and both show an ongoing interaction with their CSM years after the explosions.
They have been observed by HST Space Telescope Imaging Spectrograph (STIS) in the UV range and their spectra all show strong \mgiidoub\ lines. 
We convolved the STIS UV spectra of SN 2005ip at t=3065 days and t=4368 days \citep[data taken from][]{2020MNRAS.498..517F} with the transmission function of CSST NUV filter.
The expected magnitudes observed by CSST are $m_{NUV}$ = 20.9 and $m_{NUV}$ = 21.4, respectively.
We also did the same with the UV spectra from SN 2010jl at t=573 days \citep[data taken from][]{2014ApJ...797..118F} and the expected magnitude is  $m_{NUV}$ = 18.2.

These magnitudes are all well above the detection limit of CSST.
While neither of these two SNe closely matches our SNe II-P/L scenarios, our calculation proves that late-time \mgiidoub\ from nearby SNe can be observed by CSST, based on UV spectra gathered by HST.
\section{Observation strategy}
\label{sec:strat}
In this section we propose a set of strategies to identify ejecta-wind interaction in SNe II from CSST sky survey data.
We also discuss the expected CCSN sample size that CSST will visit during its sky survey.
\subsection{CSST survey strategy}
According to \citet{gong2019cosmology}, the CSST sky survey will cover 15000 deg$^{2}$ of area for |b| > 20$^\circ$ and 2500 deg$^{2}$ for 20$^\circ$ > |b| > 15$^\circ$.
CSST will try to visit its survey area twice during its ten-year main mission.
The detailed schedule of the survey is still being optimized, but for the purposes of this study, the exact schedule is not needed as we solely aim to give a rough estimate of the SN population CSST will encounter.

As CSST will survey 17500 deg$^{2}$ twice over ten years, this corresponds to a survey rate of ~3500 deg$^{2}$ per year, or about 8.5\% of the total sky area.

\subsection{Input SN catalog}
Considering the limiting magnitude and large FOV of the CSST sky survey, it is sure to detect many transients over its lifespan. 
However most of the detections will be one-time detections, without suitable follow-up, as the visit frequency of CSST is very low (could be years between two visits) and only very few of these detections will get a proper spectral classification.

To circumvent this problem, we propose using existing ground-based data of recent SNe as an input catalog.
Once CSST becomes operational, we can cross-check the existing SN catalog with the CSST survey images (ideally with an automated script that connects to the CSST pipeline and database API) and look for excess flux in NUV and u band.
The model spectra and light curves given in Section ~\ref{sec:methods} show that the ejecta-wind interaction becomes prominent during the late-time evolution of the SNe.
SNe that experience this interaction maintain a certain NUV brightness for many years after explosion, and possibly even longer.
Existing observations of SN2017eaw \citep[see][]{van2023disappearances} also indicate that the NUV flux can last for a few years after the SN explosion.
Therefore, this input catalog is expected to at least contain all SNe discovered in the last decade (from year 2014 and onward), with a target count on the order of $10^5$ across the entire sky.
Existing online transient platform such as Transient Name Server\footnote{https://www.wis-tns.org/} (TNS) provides sufficient and up-to-date information needed (mainly coordinates and also classifications, if available) for compiling and updating our input catalog.
Upcoming time-domain surveys such as Legacy Survey of Space and Time (LSST) will also help late-time SN studies of CSST by discovering
more SNe and recording their light curves more precisely.

With this strategy, the problem of unclassified SNe still exists as our models currently only cover typical SNe II-P/L.
If we limit our targets to spectroscopically confirmed SNe II-P/L, the overall input catalog size shrinks considerably.
However this problem is partly alleviated by the fact that SNe II-P/L are prime candidates for detecting NUV interaction signatures due to the low wind velocities of their RSG progenitors.
For other types of CCSNe, their progenitors (blue supergiants, stripped massive stars and Wolf-Rayet stars) likely have higher wind velocities, thus reducing the amount of ejecta-wind interaction power for a given mass loss rate.
Therefore most NUV bright SNe detected by CSST can be reasonably assumed to come from the ejecta-wind interactions of SNe II-P/L.
Moreover, the light curves of SNe II-P are known to exhibit a long plateau phase in V band, followed by a quick drop-off.
They are also more commonly observed in star forming galaxies. 
For SNe that have good ground based light curves, these features can be used to differentiate SNe II-P in the CSST sample from other types of SNe.
Supernovae that are bright enough to be spectroscopically classified by ground based telescopes are typically well above the detection limit of the CSST sky survey, allowing them to remain detectable for an extended period of time.

\subsection{Expected Detection}
To better estimate how many potential targets SNe II CSST will visit each year, we took a recent SN sample from the Zwicky Transient Facility (ZTF) as an example \citep[see][for details]{perley2020zwicky}. 
In their study, 226 H-rich SNe (which includes all variants of SNe II) with peak brightness over 18.5 magnitude were recorded by the ZTF over an period of 25.5 months (about 100 per year).
The m<18.5 criteria represents a good standard of completeness for spectral classification, as 93\% of events brighter than 18.5 magnitude were classified. 

Taking the 8.5\% of total sky area visited by CSST per year into account, CSST will observe about ten confirmed recent (exploded within a year) SNe II annually.
We note that the ZTF only covers region where $\delta > -30^\circ$, and infrastructure for ground-based SNe classification will likely grow over time.
When CSST becomes operational, we expect more spectroscopically confirmed SNe II than 100 per year, thus, the number of recent SNe II visited by CSST is likely to be underestimated.
Also as we mention in Section~\ref{sec:long}, the NUV signatures from interaction-powered SNe fade slowly, likely making some of these SNe visible in the NUV band for at least a few years.
However, a recent study of nearby older SNe II with the James Webb Space Telescope \citep[e.g. SN 2004et and SN 2017eaw, see][for details]{2023MNRAS.523.6048S} showed significant dust formation within the ejecta at this time scale (5-20 years after the explosions).
Dust may reduce the strength of the NUV signatures, but typically it takes a few years for dust formation to start \citep[][]{2022A&A...668A..57S}.
This leaves a time window for UV radiation largely unaffected by newly-formed dust, we expect that our t=350d and t=1000d models should fall inside this relatively dust-free window.
In the cases when dust is formed near the center of the ejecta, it may block part of the interaction powered emission.
However, the ejecta-wind interaction takes place at the outermost edge of the ejecta \citep[see Fig.2 of][]{2023A&A...675A..33D}, far above the dust-forming ejecta core.
Therefore most of the flux is only subjected to extinction outside of the SNe, such as dust newly formed in CSM and galactic as well as host extinction.

If the ejecta-wind interaction in SNe II is common and does emit most of its energy in UV, we might be able to observe many SNe with obvious NUV excess after the first few years of the CSST operation.
According to our estimate, some of the NUV-bright SNe will have ground-based spectroscopic observations during their peak brightness.
Combining the peak-time spectra and the overall light curve with the late-time NUV flux, it will be possible to constrain the mass loss history of the of SNe II-P/L progenitors.
This will offer evidence of how massive stars behave hundreds and thousands years before core collapse.

\subsection{Limitations and future works}
The mission design of CSST still poses challenges for detailed SNe observations, while some difficulties can be alleviated by combining ground-based data, others have to be dealt with novel tools or techniques developed to tackle the problems.
One of the major issues is that CSST will only scan the entire sky twice in its ten-year main mission, leading to a lack of reference image for host subtraction.
Ground-based SNe observations provide the SNe coordinates needed, but detecting a NUV/optical counterpart in the expected location of the SN is not necessarily correlated to non-fading SNe emission.

For example, companion stars and young star clusters are also possible explanations for NUV-bright sources found at the location of the SN.
Despite having an angular resolution of 0.135 arcsecond, it is still impossible for CSST to resolve single stars at Mpc distances.
At a distance of 10 Mpc, 0.135 arcsecond corresponds to a projected distance of 6.5 pc, which is about the size of star clusters.
However, the luminosity of ejecta-wind interaction ($M_{NUV}\,<\,-10$, and power input on the scale of $10^{40\sim41}$\,erg/s) is also well above the luminosity of stars that live longer than SNe II-P progenitors.
This means that it is possible to rule out stellar contamination in most cases, as the more massive, NUV-bright stars born together with SNe II-P progenitors have already exploded.

In future works, it might be also possible to develop a model or algorithm to separate nebular emission of late-time SNe from stellar emission based on color from all seven bands of CSST.
Otherwise, we can at least rely on the difference between various sets of images gathered by CSST as it will survey the sky twice during its main mission.
This can be used to identify fading sources (i.e., SNe) from constant sources (i.e., star clusters or background and foreground stars).

Host and galactic extinction may also affect the detection of late-time SNe emission, especially towards the shorter wavelengths.
For our study, the total amount of host and galactic extinction can be pre-determined or estimated from the peak-time color of the SNe, and we can use this information to apply corrections to late-time observations of CSST.
Also, CSST will mainly survey outside of the galactic plane, therefore, galactic extinction is unlikely to have a great impact.
\section{Conclusions}
\label{sec:conclusion}
In this paper, we demonstrated the capability of CSST to detect ejecta-wind interaction signatures of SNe II-P via the NUV sky survey. 
The ejecta-wind interaction is caused by the SN ejecta colliding with wind material previously expelled by steady wind-mass loss of the progenitor massive star. 
As wind-mass loss is common in massive stars, in particular, during the post main sequence stages (e.g., the RSG phase of stars with initial mass of $\mathrm{}10-25\,M_{\sun}$), we expect this interaction to be universal in SNe II-P.
However, the energy released by this interaction mainly appears as UV radiation, which is inaccessible for ground-based telescopes.

We convolved the modeled spectra from \citet{dessart2022modeling} and \citet{2023A&A...675A..33D} with the filter transmission function of CSST to obtain light curves of different input models as observed by CSST. 
The boxy \mgiidoub\ feature of the spectra contributes significantly to the flux received by the NUV band, resulting in plateau-like NUV light curves for models affected by the ejecta-wind interaction. 
This NUV flux is detectable by CSST out to 100$\sim$400\,Mpc, depending on the magnitude of the interaction power.

We propose using CSST to identify such NUV-excess SNe by cross-checking previous SNe discoveries with the sky survey NUV imagery.
In order to be sure that the SNe observed by CSST are indeed SNe II, we can limit the input SNe to those with ground-based spectra around their peak brightness.
This reduces size of input SNe II catalog to $\sim$100 per year, but CSST will still be able to observe about ten of such SNe II annually, which had exploded within that year.
After a few years of operation, CSST is expected to provide several tens of NUV observations for nearby SNe II.

If the ejecta-wind interaction powered NUV signature predicted in this work turns out to be common among SNe II, it can potentially become a novel tool for inferring the mass loss history and for constraining the evolution of the SN progenitor. 

\begin{acknowledgements}
     We thank the anonymous referee and our editors for their careful reading of the paper which improves its quality and clartiy. We are grateful to Xiangcun Meng, Ningchen Sun, Yi Yang, Lifu Zhang, Qiyuan Cheng, Hao Shen for their fruitful discussions. This work is supported by the National Natural Science Foundation of China (NSFC, Nos.\ 12125303, 12288102, 12090040/1, 11873016), the National Key R\&D Program of China (Nos.\ 2021YFA1600401 and 2021YFA1600400), the Chinese Academy of Sciences (CAS), the International Centre of Supernovae, Yunnan Key Laboratory (No. 202302AN360001), the Yunnan Fundamental Research Projects (grant Nos.\ 202201BC070003, 202001AW070007) and the ``Yunnan Revitalization Talent Support Program''—Science \& Technology Champion Project (No.~202305AB350003). We also acknowledge the science research grants from the China Manned Space Project with No. CMS-CSST-2021-A10.
\end{acknowledgements}

%-------------------------------------------------------------------

\bibliographystyle{aa} % style aa.bst
\bibliography{aa} % your references Yourfile.bib

\begin{appendix}
\section{Additional figures}

\begin{figure}
	% To include a figure from a file named example.*
	% Allowable file formats are eps or ps if compiling using latex
	% or pdf, png, jpg if compiling using pdflatex
	\includegraphics[width=\columnwidth]{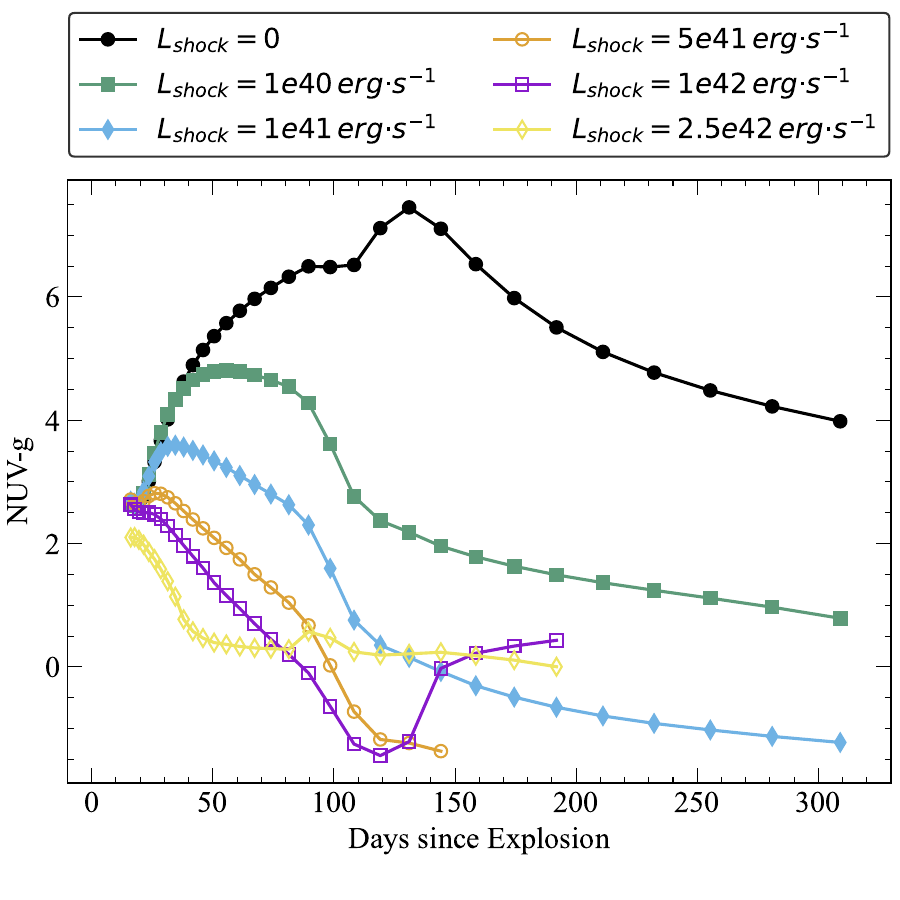}
    \caption{Same as Fig.~\ref{fig:CSST_color} of main-text, but for NUV-g color.}
    \label{fig:CSST_color-g}
\end{figure}

\begin{figure}
	% To include a figure from a file named example.*
	% Allowable file formats are eps or ps if compiling using latex
	% or pdf, png, jpg if compiling using pdflatex
	\includegraphics[width=\columnwidth]{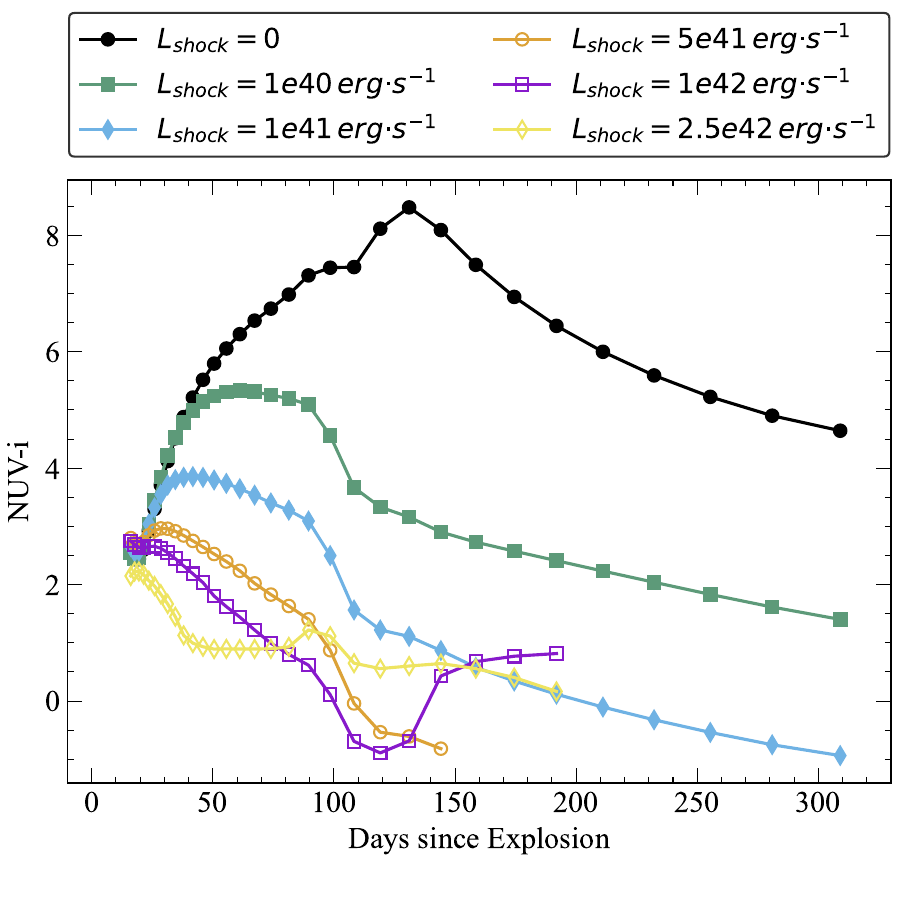}
    \caption{Same as Fig.~\ref{fig:CSST_color} of main-text, but for NUV-i color.}
    \label{fig:CSST_color-i}
\end{figure}
\end{appendix}

\end{document}